\documentstyle[12pt]{article}

\begin{document}

\def\lsim{\ \matrix{<\cr\noalign{\vskip-7pt}\sim\cr} \ }
\def\gsim{\ \matrix{>\cr\noalign{\vskip-7pt}\sim\cr} \ }

\begin{titlepage}

\hfill{PURD-TH-93-06}

\hfill{SISSA-/93/49-A}

\hfill{DFPD/TH/93/27}

\vskip 1cm

\centerline{\Large \bf Baryogenesis, Domain Walls and the Role of Gravity}

\vskip 1.5cm

\centerline{{\bf Henry Lew$^{(a)}$}\footnote{email: lew\%purdd.hepnet@LBL.Gov}
{\bf and Antonio Riotto$^{(b)(c)}$}\footnote{email: riotto@tsmi19.sissa.it}}

\vskip 1.5cm

\noindent
\centerline{{\it (a) Physics Department, Purdue University, West Lafayette,
IN 47907-1396, U.S.A.}}

\vskip 2 mm
\noindent
\centerline{{\it (b) International School for Advanced Studies, SISSA,
via Beirut 2-4, I-34014 Trieste, Italy.}}

\vskip 2mm
\noindent
\centerline{{\it (c) Istituto Nazionale di Fisica Nucleare, Sezione di Padova,
I-35100 Padua, Italy.}}
\vskip 1.5 cm

\centerline{\large\bf Abstract}
\vskip 1cm
\noindent
\baselineskip 20pt
It has been recently speculated that global symmetries are
broken by gravity. We propose a scenario for the generation
of the baryon asymmetry in the early Universe in which the
domain walls predicted by theories with discrete symmetries
become unstable due to these Planck scale effects. The relative
motion of the decaying walls can provide a mechanism for the departure
from thermal equilibrium necessary to have a nonvanishing baryon
asymmetry. In particular, we implement this idea within the
frameworks of the Left-Right and Quark-Lepton symmetric models.
\end{titlepage}

\leftline{\bf 1. Introduction}
\vskip 5mm
\baselineskip 20pt
With the emergence of a gauge theory of the strong and
electroweak interactions of particle physics and the
standard hot big bang model of cosmology it became possible
to attempt to answer the fundamental question of why there
exists a matter-antimatter asymmetry in the Universe.
Over the years there have been a number of scenarios proposed to
describe the baryon asymmetry of the Universe (BAU) \cite{revs}.
Recent attempts include baryogenesis at the electroweak scale
using just the standard model (SM) of particle physics \cite{ewb}.
Though this scenario is attractive, it nevertheless requires
the anomalous interactions to be out of equilibrium after the
completion of the electroweak phase transition (EWPT), otherwise
the BAU generated at the Fermi scale would be washed out by
sphalerons \cite{ewb}. This requirement translates into
an upper bound on the mass of the scalar Higgs particle
already excluded by the present LEP lower bound on the
Higgs mass \cite{hb}. Moreover, for such a scenario to work, the
EWPT must be of the first order and proceed via bubble nucleation,
whereas recent analysis seem to indicate that it is very weakly
first order or even second order \cite{z}.

In this paper we propose a new mechanism to generate the BAU
in two particle physics models which predict new physics
beyond the SM. The two models that we will consider are the
Left-Right (LR) \cite{LR} and the Quark-Lepton (q-$\ell$) \cite{ql}
symmetric models. Both of these models will have interesting
phenomenology which hopefully can be tested in upcoming
experiments in the near future.

The basic requirements for the generation of the BAU \cite{sak} are:
(1) baryon number violation, (2) $C$ and $CP$ violation and
(3) a departure from thermal equilibrium. The key ingredients
in our proposed mechanism which are needed to satisfy these
basic requirements are given by (i) the lepton number ($L$) violating
interactions of the right-handed neutrino, $\nu_R$, (ii) domain walls
and (iii) the intrinsic $C$ and $CP$ violations of the models under
consideration. The right-handed neutrinos are responsible for
establishing an initial lepton number asymmetry. The lepton number
asymmetry is then converted to a baryon number ($B$) asymmetry via the
sphaleron at electroweak energies. Since the LR and q-$\ell$
models have discrete symmetries which are spontaneously broken,
the formation of domain walls will result \cite{vilenkin}. Percolation
theory seems to indicate that an ``infinite'' domain corresponding to
each vacuum will form and therefore an ``infinite'' domain wall of
complicated topology will appear \cite{per}.

Stable domain walls pose a problem in the standard big bang scenario
unless the discrete symmetry breaking scale is very low \cite{zel}.
One way to resolve this problem is to break the discrete symmetry
explicitly and thereby making the domain wall unstable.
Recently there has been speculation that discrete symmetries
which are not gauged are broken by gravity and that even very
tiny effects of gravity may suffice to make the domain walls
cosmologically harmless \cite{grav}.
We propose a scenario in which the discrete symmetry is broken by
gravity so that the domain walls become unstable and eventually
disappear. This will provide a mechanism for the departure from
thermal equilibrium.  We wish to stress that in our scenario
the EWPT is not required to take place via bubble nucleation but
might be harmlessly of the second order.

We first describe the details our mechanism for baryogenesis
in the LR symmetric model \cite{Holdom} and then in the q-$\ell$
symmetric model.

\vskip 1cm
\leftline{\bf 2. LR symmetic model}
\vskip 5mm

The LR symmetric model of weak interactions is based
on the gauge group
\begin{equation}
SU(2)_L \otimes SU(2)_R \otimes U(1)_{B-L}
\end{equation}
supplemented by a discrete symmetry between the left and right
sectors of the model. The fermion content of the model
is given by
\begin{eqnarray}
F_L &\sim & (2,1)(-1),\ \ F_R \sim (1,2)(-1), \nonumber \\
Q_L &\sim & (2,1)(1/3),\ \ Q_R \sim (1,2)(1/3),
\end{eqnarray}
where $F$ and $Q$ denote the leptons and quarks respectively.
The Higgs sector is given by
\begin{equation}
\Delta_L \sim  (3,1)(2),\ \ \Delta_R \sim (1,3)(2), \nonumber
\end{equation}
\begin{equation}
\Phi \sim  (2,2)(0).
\end{equation}
The symmetry breaking goes as follows when the Higgs fields
acquire nonzero vacuum expectation values (VEVs):
\begin{eqnarray}
&SU(2)_L \otimes SU(2)_R  \otimes U(1)_{B-L} &\nonumber \\
&\langle\Delta_L\rangle = 0 \ \downarrow\
\langle\Delta_R \rangle \not= 0 &\nonumber \\
&SU(2)_L \otimes U(1)_Y&\nonumber \\
&\ \ \ \ \ \ \ \ \downarrow\
\langle\Phi\rangle \not= 0 &\nonumber \\
& U(1)_Q&
\end{eqnarray}
where $Y = 2I_{3R} + (B-L)$ and $Q = I_{3L} + Y/2$.

As the temperature in the early Universe drops to about
$T \sim \langle \Delta_R \rangle$ a domain wall forms due
to the breaking of the left-right symmetry $\Delta_{R}
\leftrightarrow \Delta_{L}$ \cite{domlr}.
To avoid the problems associated with stable domain walls we
will assume that gravity explicitly breaks the discrete LR symmetry.
This is not as ad hoc as it sounds since we know the gravitational
interaction exists and presumably when a satisfactory quantum theory
of gravity becomes available then the explicit breaking becomes
calculable in principle.

The domain walls separate the two different domains which were
formed independently after the cosmological phase transition
when the discrete symmetry was spontaneously broken. Just after the
formation of the wall, there exists massive left-handed neutrinos in one
domain while in the other domain the massive neutrinos are
right-handed. Assuming that gravity lifts the vacuum degeneracy
of the two domains then the true vacuum will expand while the false
one shrinks. If we take the true vacuum to be the domain with the
massive right-handed neutrinos then an excess of massive right-handed
neutrinos will result as those originally massless right-handed neutrinos
move from the false vacuum to the true one. This is the qualitative
picture of our out-of-equilibrium scenario. We now proceed to give
some quantitative details.

Consider the two domains, denoted by L and R, on either side of the
domain wall. They are the local minima of the Higgs potential which
have their degeneracy broken by the following terms induced by gravity
\begin{equation}
{K \over M_p^2} \left( \Delta_L^\dagger \Delta_L \right)^3
+ {J \over M_p^2} \left( \Delta_R^\dagger \Delta_R \right)^3\: + \:
\hbox{higher order terms},
\label{break}
\end{equation}
where $K \not= J$ and $M_p$ is the Planck scale.
These terms do not prevent the formation of the domain walls
at $T\sim \langle\Delta_{R}\rangle$ since the energy difference
between the two almost degenerate vacua is of the order of
$\alpha\langle\Delta_{R}\rangle^4$,
where $\alpha\sim \langle\Delta_{R}\rangle^2/M_{p}^2$,
and $\alpha\ll 0.2\:\beta$, $\beta$ indicating the generic
coefficient of the quartic terms in the
potential $V\left(\langle\Delta_{R}\rangle,
\langle\Delta_{L}\rangle\right)$ \cite{gra}.

As a result of the presence of the terms in Eq. (\ref{break})
the wall moves and the region of false vacuum, call it L, shrinks.
The energy released by the decay of the wall is
\begin{equation}
\Delta\rho \sim |K-J| {\langle \Delta_R \rangle^6 \over M_p^2},
\end{equation}
where we will assume that $|K-J|$ is of order one.
The domain wall decays in the time \cite{sik}
\begin{equation}
t_W \sim {\eta \over \Delta\rho}
\sim {M_p^2 \over  \langle \Delta_R \rangle^3},
\label{tw}
\end{equation}
where $\eta \sim \langle \Delta_R \rangle^3$ is the mass per surface
area of the wall. By requiring that the wall decays before the
EWPT so that the anomalous $B$-violating
interactions are still active gives
\begin{equation}
\langle \Delta_R \rangle > \left(M_p T_{EW}^2 \right)^{1\over 3}
\sim 10^8 \hbox{ GeV},
\end{equation}
where $T_{EW}$ is the EWPT temperature.
(We also need to check that the domain walls disappear before
a time $t_0 = M_p^2/\eta$ so that the energy density and the
cosmic microwave background radiation of the observed Universe
are not significantly affected \cite{zel}. By requiring that
$t_W \lsim t_0$ gives $\langle \Delta_R \rangle^3 \gsim \eta$.)

During this time the right-handed neutrino gets a Majorana mass,
$m_{\nu_R} \sim \langle \Delta_R \rangle$, from the Lagrangian
\begin{equation}
{\cal L}_{Yuk} =
h\left[\overline{F_L} (F_L)^c \Delta_L
+ \overline{F_R} (F_R)^c \Delta_R \right] + {\rm H.c.}.
\end{equation}
The number density of the out-of-equilibrium right-handed
neutrinos which are produced and consequently decay is of the
order of
\begin{equation}
n_{\nu_{R}}\sim {\Delta\rho \over m_{\nu_R}}
\sim {\langle \Delta_R \rangle^5 \over M_p^2}.
\label{nnu}
\end{equation}
Note that the time by which the domain walls disappear
(see Eq. (\ref{tw})) corresponds to a temperature of
\begin{equation}
T \simeq \sqrt{M_p \over t_W}
= \langle \Delta_R \rangle
\sqrt{\langle \Delta_R \rangle \over M_p}.
\label{Tdis}
\end{equation}
For $\langle \Delta_R \rangle \sim 10^8$ GeV and $M_p \sim 10^{19}$ GeV,
this temperature is about 300 GeV. Also note that this temperature is less
than the temperature of the right-handed neutrinos,
$T_{\nu_R} \sim \langle \Delta_R \rangle$, and hence we expect
them to be away from their equilibrium distributions.
This means that the out-of-equilibrium conditions due to the
moving wall persists until about 300 GeV. Below this temperature
equilibrium is re-established and $L$ is converted to $B$ via
sphaleron processes which conserves $\left(B-L\right)$ \cite{ewb}.

The BAU can now be calculated from the lepton number asymmetry
due to the decays of the right-handed neutrino, {\it i.e.}
\begin{eqnarray}
\nu_{R_{i}} & \rightarrow & F_{L_{j}} + \bar\Phi \nonumber \\
\nu_{R_{i}} & \rightarrow & \bar F_{L_{j}} + \Phi.
\end{eqnarray}
By using Eq. (\ref{nnu}) the lepton number density is given by
\begin{equation}
n_{L_{i}} - n_{\bar {L}_{i}}
= {\langle \Delta_R \rangle^5 \over M_p^2} \epsilon_{i}
\end{equation}
where $\epsilon_{i}$ is the difference between particle-antiparticle
branching ratios \cite{FY}
\begin{equation}
\epsilon_{i}=\frac{1}{2\pi\left(\lambda\lambda^{\dag}\right)_{ii}}
\sum_{j}\left({\rm Im}\left[\left(\lambda\lambda^{\dag}\right)_{ij}
\right]^2\right) f\left(m_{\nu_{R_{j}}}^2/m_{\nu_{R_{i}}}^2\right),
\end{equation}
where
\begin{equation}
f(x)=\sqrt{x}\left[1-\left(1+x\right){\rm ln}\left(\frac{1+x}{x}\right)
\right],
\end{equation}
and the $\lambda$'s denote the Yukawa couplings between right-handed
and left-handed neutrinos through the Higgs bi-doublet $\Phi$ and are
assumed to be complex to give a source of CP violation.

At $T \sim T_{EW}$ the lepton number produced by right-handed neutrino
decays is
\begin{eqnarray}
L & = & {n_L - n_{\bar L} \over s } \nonumber \\
  & \sim & {\langle \Delta_R \rangle^5 \over 10^2\: T_{EW}^3 \:M_p^2 }\:
\epsilon,
\end{eqnarray}
where the entropy density, $s$, is given by
$s \sim g_{*s}T^3 \sim 10^2\: T^3$. Since the domain walls
decay just before dominating the energy density of the Universe,
$L$ is not overdiluted by the energy released during the
domain wall decay.

A nonzero baryon number is then induced via the $(B-L)$
conserving sphaleron processes given by
\begin{eqnarray}
B & = & \kappa\: L \nonumber \\
  & \sim & \kappa\:
{\langle \Delta_R \rangle^5 \over 10^2\: T_{EW}^3 \:M_p^2 }\:\epsilon,
\label{bau}
\end{eqnarray}
where $\kappa$ is a numerical factor of ${\cal O}(1)$ and
can be easily calculated from Ref. \cite{har}.
Note that at $T_{EW}\ll \langle\Delta_{R}\rangle$, $SU(2)_{R}$ sphalerons
are no longer active and thus they do not affect Eq.
(\ref{bau})\footnote{$SU(2)_{R}$ sphalerons are crucial for the
mechanism described in Ref. \cite{right} where right-handed neutrinos
are suppose to scatter off first-order phase transition bubbles in the
framework of LR models.}.

For $\langle \Delta_R \rangle  \sim 10^8$ GeV,
$T_{EW} \sim 10^2$ GeV and $M_p \sim 10^{19}$ GeV gives a baryon
asymmetry of about $B \sim 10^{-6} \epsilon$. Since the
present baryon asymmetry lies in the range
$\left(4-5.7\right)\times 10^{-11}$ \cite{ol},
$\epsilon$ has to be as large as $10^{-5}$, which can be obtained
for reasonable values of the $\lambda$'s, e.g.
$\lambda \sim {\cal O}(10^{-2})$.

We also need to check that the initial baryon asymmetry generated
by the domain walls and the decaying right-handed neutrinos is
not erased by a combination of other lepton violating interactions
and sphaleron processes \cite{FY2} at temperatures below that
corresponding to the disappearance of the domain walls
(see Eq. (\ref{Tdis})). The lowest dimension $L$-violating
operator in the effective low energy theory is given by
\begin{equation}
{\cal L}_{\Delta L = 2}
= {m_{\nu_L}\over v^2}F_L F_L \Phi\Phi + {\rm H.c.},
\end{equation}
where $m_{\nu_L}$ is the mass of the left-handed neutrino
and $v$ is the electroweak breaking scale. The interaction
rate of these $\Delta L = 2$ processes is
$\Gamma_{\Delta L = 2}
\simeq m_{\nu_L}^2 T^3/ \left(\pi^3 v^4 \right)$. For the
survival of the pre-existing asymmetry we require this interaction
rate to be less than the expansion rate of the Universe,
{\it i.e.}, $H \sim T^2/ M_p$, where $T$ is given by Eq. (\ref{Tdis}).
This results in a bound on the $SU(2)_R$ breaking scale
\begin{equation}
\langle\Delta_R \rangle
\gsim {\lambda^8 \over \pi^6 h^4} \:M_{p}.
\end{equation}
where $\lambda$ is the Yukawa coupling for the Dirac mass
term and $h$ is the one for the Majorana mass.
So for conservative values of the couplings, e.g. $\lambda \sim 10^{-2}$
and $h \sim 10^{-1}$, the bound on  $\langle\Delta_R \rangle$ can
easily be made lower than $10^8$ GeV, consistent with the bound
from the lifetime of the domain wall.

\vskip 1cm
\leftline{\bf 3. The q-$\ell$ model}
\vskip 5mm

The analysis for generating the BAU in q-$\ell$ model is
essentially the same as that in LR model. In the following
we outline how the mechanism is applied to the q-$\ell$ model.

Consider the q-$\ell$ model described by the gauge group
\begin{equation}
SU(3)_{\ell} \otimes SU(3)_q \otimes SU(2)_L \otimes U(1)_X.
\end{equation}
SU(3)$_q$ is the usual colour group and SU(3)$_{\ell}$
is its leptonic partner. In addition there is a discrete $Z_2$
symmetry between the quarks and leptons. The fermion content of
the model is given by
\begin{eqnarray}
F_L &\sim & (3,1,2)(-1/3),\ \ E_R \sim (3,1,1)(-4/3),\ \ N_R \sim
(3,1,1)(2/3), \nonumber \\
Q_L &\sim &(1,3,2)(1/3),\ \ u_R \sim (1,3,1)(4/3),\ \ d_R \sim
(1,3,1)(-2/3).
\end{eqnarray}
The Higgs sector is given by\footnote{In the minimal q-$\ell$
model the Yukawa Lagrangian yields the tree-level mass matrix
relations, $M_u = M_e$ and $M_d = M_\nu^{\hbox{Dirac}}$, where
$u$ refers to the charge 2/3 quarks, $e$ refers to the charged
leptons etc. The latter mass relation is broken by the Majorana
mass terms for the neutrinos. The former relation can be evaded,
for example, by introducing an additional Higgs doublet.}
\begin{eqnarray}
&\chi_1 \sim (\overline{3},1,1)(-2/3),\quad
\chi_2 \sim (1,\overline{3},1)(2/3), & \nonumber \\
&\Delta_1 \sim (6,1,1)(4/3), \quad
\Delta_2 \sim (1,6,1)(-4/3), & \nonumber \\
& \phi \sim (1,1,2)(1). &
\end{eqnarray}
The symmetry breaking pattern can be summarised as follows:
\begin{eqnarray}
&SU(3)_{\ell} \otimes SU(3)_q  \otimes  SU(2)_L
\otimes U(1)_X &\nonumber \\
&\langle\Delta_1\rangle\ \downarrow\
\langle\chi_1 \rangle &\nonumber \\
&SU(2)' \otimes SU(3)_q \otimes SU(2)_L \otimes U(1)_Y&\nonumber \\
&\ \ \ \ \ \ \downarrow\
\langle\phi\rangle &\nonumber \\
&SU(2)' \otimes SU(3)_q \otimes U(1)_Q&
\end{eqnarray}
The SU(2)$'$ is an unbroken gauge symmetry. This gauge force
is expected to be asymptotically free. In analogy with QCD,
we assume that it confines all the exotic SU(2)$'$ coloured states,
so that at large distances only colour singlet states exist
in the spectrum.

The Higgs scalars, $\Delta_{1,2}$, play an analogous role to
that of $\Delta_{L,R}$ in the LR model. The right-handed neutrinos
gain Majorana masses through the term
\begin{equation}
{\cal L}_{Yuk} =
h\left[\overline{N_R} (N_R)^c \Delta_1
+ \overline{d_R} (d_R)^c \Delta_2 \right] + {\rm H.c.},
\end{equation}
when $\Delta_1$ develops a nonzero VEV while the VEV of
$\Delta_2$ remains zero. At the same time the lepton number
symmetry is broken so that an initial lepton number asymmetry
can be generated. Without $\Delta_{1,2}$, there will exist
a global lepton number symmetry, unbroken at tree-level, even
though the q-$\ell$ symmetry will be broken by
$\langle\chi_1 \rangle$. Note that even if
$\langle\Delta_1\rangle \sim 10^8$ GeV, new physics in the form
of exotic ``leptons'' may still be observable in the 100 GeV
to 10 TeV region provided
$\langle\chi_1 \rangle \ll \langle\Delta_1\rangle$.

The domain wall picture in the q-$\ell$ model \cite{domql} is
somewhat different from that of the LR model. Consider the
situation for the q-$\ell$ model just after wall formation.
The domains on each side of the wall are described by
\begin{eqnarray}
\hbox{region 1} & : & \langle\Delta_1 \rangle \not= 0,
\quad \langle \Delta_2 \rangle = 0 \nonumber \\
\hbox{region 2} & : & \langle\Delta_1 \rangle = 0,
\quad \langle\Delta_2 \rangle \not= 0.
\end{eqnarray}
The particles in region 1 will be the charged-conjugated
versions of those in region 2. When the massive right-handed
neutrino passes the wall from region 1 to region 2 it becomes
a massless right-handed d-quark with hypercharge $Y= -2/3$ and
vice versa. Similarly a massive right-handed neutrino passing
the wall from region 2 to region 1 will become a massless right-handed
(anti-)d-quark with $Y= 2/3$. When the discrete symmetry is broken
by gravitational effects, so that $\langle\Delta_1 \rangle \not=
\langle\Delta_2 \rangle$, the wall will move since one domain will
shrink at the expense of the other. There will be an excess of right-handed
neutrinos created from the conversion of right-handed (massless)
d-quarks as they move from the false vacuum to the true one.
The remaining scenario for baryogenesis then proceeds like that in
the LR model.

\vskip 1cm
\leftline{\bf 4. Conclusions}
\vskip 5mm
In the present paper we have shown that the role of gravity
might be crucial not only for rendering topological defects like
domain walls cosmologically harmless, but also for generating the baryon
asymmetry in the early Universe. The point is that Planck scale effects
{\it could} introduce new interaction terms which explicitly break the
global discrete symmetries of the Lagrangian at tree level. If this is
the case, then domain walls, which appear when discrete symmetries are
spontaneously broken, can decay at a temperature below the typical scale
of the theory, providing the out-of-equilibrium condition necessary to
have a nonvanishing $B$. This mechanism can be successfully applied to the
LR and q-$\ell$ symmetric models if the R and the q-$\ell$ breaking scales
are of the order of $10^{8}$ GeV respectively. Moreover, in our scenario,
the EWPT need not be first order.

\vskip 1cm
\centerline{\bf Acknowledgements}
\vskip 1mm
\noindent
It is a pleasure to express our gratitude to P. Catelan, G. Dvali,
S. Matarrese and G. Senjanovi\'c for stimulating and enlightening
discussions. This work was supported in part by a grant from the DOE.

\end{document}